\documentclass[12pt]{revtex4}
\begin{document}
\newcommand\al{\alpha}
\newcommand\psial{\psi_{\al}}
\newcommand\wal{w_{\al}}
\newcommand\Cal{C_{\al}}
\newcommand\xd{\dot x}\def\qd{\dot q}
\newcommand\xin{x_i}
\newcommand\xfi{x_f}
\newcommand\xp{x^{\prime}}
\newcommand\xpi{x_i^{\prime}}
\newcommand\xpf{x_f^{\prime}}
\def\tp{\tau^\prime}
\newcommand\s{\sigma}
\newcommand\xb{\bar x}
%\hfill {KFKI-RMKI-23-MARCH-1993}
\title{UNIQUE FAMILY OF CONSISTENT HISTORIES IN THE CALDEIRA-LEGGETT MODEL
\footnote{p15-24 in: Stochastic Evolution of Quantum States in Open Systems and in Measurement
Processes, eds.: L.Di\'osi and B.Luk\'acs (World Scientific, Singapore, 1994)}}
\author{Lajos Di\'osi}\email{diosi@rmki.kfki.hu}
\affiliation{
KFKI Research Institute for Particle and Nuclear Physics\\
             H-1525 Budapest 114, POB 49, Hungary}
\begin{abstract}
Based on the standard statistical interpretation of mixed quantum
states,
a unique family of consistent histories has been constructed for
the quantum Brownian motion in the Caldeira-Leggett reservoir.
Analytic solutions have been shown in the Markovian regime:
they are uniquely defined coherent wave packets travelling near
classical trajectories.
\end{abstract}
\maketitle

\section{Introduction}

Since the famous work of von Neumann [1], the statistical operator
$\rho$ has been used to represent
the general quantum state of a given
quantum system. The statistical interpretation of $\rho$ is
given in frames of the quantum measurement theory.

For the recent years, serious efforts have been made to derive
statistical interpretation from the (slightly modified)
quantum dynamics itself [2]. To achieve a similar goal,
other authors have proposed a certain history-formulation [3]
of the quantum mechanics instead of the ordinary one.
We do not intend to discuss any of the preceeding
proposals. Rather we are going to propose a unique and exact
history-interpretation for a particular quantum system.
Although we use the terminology of works [3], we would express
our results in the conservative (standard) language of the
quantum mechanics as well (cf.Ref.[4]). Lessons learned from
the works [2] are essential even if not made explicit in the
present paper.

If one ignores the measurement theory,
it is still possible to infer a certain
statistical content from a general state $\rho$. We can always
decompose $\rho$ into the weighted sum of pure
state statistical operators:
$$
\rho = \sum_{\al}\wal\psial\psial^{\dagger}.\eqno(1.1)
$$
Accordingly, we can interprete the given (mixed) quantum state
$\rho$ as follows: the
state of the system is just a pure state singled out at random
from the set $\{\psial\}$, with probablities $\{\wal\}$,
respectively. Hence, for mixed quantum states a genuine
statistical interpretation is possible, even without referring to
the concept of quantum measurement.

The pure states $\psial$ may be called {\it consistent states}
because the above statistical interpretation is fully consistent
with what is expected of a usual statistical ensemble,
cf.Refs.[3].

For the consistent states $\{\psial\}$ we can introduce
the notion of {\it decoherence} [3]. Instead of strict decoherence,
usually we find a weaker one, e.g. an asymptotic decoherence:
$$
\psi_{\al_1}^\dagger\psi_{\al_2}\rightarrow\ 0\eqno(1.2a)
$$
if labels $\al_1$ and $\al_2$ become "very" different.
To be precise, assume the existence of Euclidean norm
on the space of labels. Thus the condition for the limes (1.2a)
reads:
$$
\Vert\al_1-\al_2\Vert\rightarrow\ \infty.\eqno(1.2b)
$$
We wish to emphasize that, at least in our work, decoherence is
not a logical necessity to assign consistent probabilities
to the terms
of the decomposition (1.1) of mixed states. No doubt,
asymptotic decoherence (1.2ab) shows up as a characteristic
feature of our consistent states, as will be seen in Par.III.

The decomposition (1.1) is trivial and unique if the state $\rho$
is already pure. Pure states has no classical statistical
content independent of further assumptions like, e.g.,
performing measurements on the pure state. In general, a pure
state turns to be mixed provided we ignore variables
belonging to a certain factor space of the
system's Hilbert space. Technically, one has to take a trace of
the original statistical operator $\rho$ over the factor space
of the ignored variables. In recent works this has been termed
as {\it coarse graining} [3].
Coarse graining can thus produce mixed
quantum states which, in turn, will possess classical statistical
content in terms of consistent states and their probabilities,
as is seen from Eq.(1.1).

For mixed states, unfortunately, the decomposition (1.1) is
not unique. We shall
propose a certain way to obtain unique results in case of a
wellknown model of coarse graining. The proposal relies upon
the fact that certain decompositions are distinguished by
the dynamics of the coarse grained system.

\section{Consistent histories in coarse grained systems}

For a given coarse grained (i.e. reduced) dynamics,
the statistical
operator satisfies {\it linear} evolution equation of the general
form
$$
\rho(t)=J(t)\rho(0),\ \ t>0,\eqno(2.1)
$$
where $J$ is the evolution {\it superoperator}. A basic
feature of $J$ is that Eq.(2.1) generates mixed states
from pure ones permanently. The formal generalization of
Eq.(1.1) reads:
$$
\rho(t) = \sum_{\al}\wal(t)\psial(t)\psial^{\dagger}(t).\eqno(2.2)
$$

This equation must be conform with the Eq.(2.1), regarding
especially the linearity of the superoperator $J$. This
condition is easy to meet if the unnormalized states
$\sqrt{\wal(t)}\psial(t)$, too, satisfy linear
evolution equations. Hence, we shall assume that
$$
\psial(t)={1\over\sqrt{\wal(t)}}\Cal(t)\psial(0)\eqno(2.3)
$$
where $\Cal(t)$ is time dependent linear evolution operator for
the consistent state $\psial(t)$. Observe that the normalized
state $\psial(t)$ satisfies nonlinear equation though the
nonlinearity is only caused by the normalizing prefactor. It
is fixed just by the normalization condition:
$$
\wal(t)=\Vert\Cal(t)\psial(0)\Vert^2.\eqno(2.4)
$$

Let us substitute Eq.(2.3) into Eq.(2.2) and compare
the result with Eq.(2.1). Then, the evolution superoperator can
be written in terms of the evolution operators of the consistent
states:
$$
J(t)=\sum_{\al}\Cal(t)\otimes\Cal^\dagger(t).\eqno(2.5)
$$
This superoperator is, as expected, linear.

Let us summarize our proposal. Assume a coarse grained system
is given, with known linear evolution superoperator $J$ in Eq.(2.1).
Single out a certain
dyadic decomposition (2.5) of $J$ in terms of linear evolution
operators
$\Cal$. Once the operators $\Cal$ have been specified, the
coarse grained dynamics can be described in terms of
{\it consistent histories} $\{\psial(\tau),\tau\in[0,t]\}$
generated by the history operators $\Cal$ via the
nonlinear evolution equation
(2.3). A given history is realized with probability $\wal(t)$
as expressed by Eq.(2.4).

It is most important to realize that the operators $\Cal$ determine
the dynamics as well as the statistics of the consistent histories.
Still the choice of the $\Cal$ is not unique since Eq.(2.5) offers
little constraint on it.

\section{Consistent histories in the Caldeira-Leggett model}

Nontrivial, i.e. nonunitary evolution equations of type (2.1) are
usually not easy to derive. A wellknown exception, calculable
explicitly, is the evolution of the state $\rho$ of a Brownian
particle interacting with a given bosonic reservoir being in
thermal equilibrium [5]. Coarse graining is meant by tracing out
the reservoir variables. Then standard calculations lead to the
exact form of the evolution superoperator:
$$
J(\xfi,\xpf,\xin,\xpi,t)\eqno(3.1)
$$
as expressed by Eqs.(A.6) and (A.7) in coordinate representation.

Following the proposal of Par.II, the evolution superoperator (3.1)
will be decomposed into a specific dyadic form (2.5).
We shall exploit the fact that $J$ is
expressed by double Gaussian integrals over the {\it paths}
$x(\tau)$ and $\xp(\tau),\ \tau\in[0,t]$.
Also the label $\alpha$ of the operators $\Cal$
will actually be a path $\xb(\tau)$
rather than a number and, consequently, the summation
in Eq.(2.5) will be replaced by functional integration
over $\xb$. In coordinate representation one writes
Eq.(2.5) in the form:
$$
J(\xfi,\xpf,\xin,\xpi,t)=
\int D\xb C_{[\xb]}(\xfi,\xin,t)C_{[\xb]}^\ast(\xpf,\xpi,t).\eqno(3.2)
$$
If we choose Gaussian form for the operator kernel
$C_{[\xb]}(\xfi,\xin,t)$ then the superoperator functional $J$, too,
will be Gaussian. By a clever choice, we can just obtain
the required form (A.6).

Let us assume the following Gaussian expression for the history
operators:
$$
C_{[\xb]}(\xfi,\xin,t)=\int Dx exp\left({i\over\hbar}S[x]\right)
                      \Phi_{[\xb]}[x]\eqno(3.3a)
$$
where
\begin{eqnarray}~~~~~~~~~~
\Phi_{[\xb]}[x]=exp\biggl(
&&\!\!\!\!\!\!
-{2i\over\hbar}\int_0^t d\tau\int_0^\tau d\s x(\tau)\eta(\tau-\s)\xb(\s)\nonumber\\
&&\!\!\!\!\!\!
-{1\over\hbar}\int_0^t d\tau\int_0^t d\s
           [x(\tau)-\xb(\tau)]\tilde\nu (\tau-\s)[x(\s)-\xb(\s)]
           \biggr)~~~~~~~~~~~~~~~~(3.3b)\nonumber\end{eqnarray}
and $\tilde\nu$ is a certain modification of the noise kernel (A.9a),
specified below. Let us substitute Eqs.(3.3ab) into Eq.(3.2)
and perform the Gaussian functional integration over $\xb$. The
resulting expression will coincide with the form given by Eqs.(A.6)
and (A.7),
provided the following constraint fulfills [cf.Eq.(3.11) of Ref.6]:
$$
\nu = \tilde\nu + \eta^r\tilde\nu^{-1}\eta^a\eqno(3.4)
$$
where we applied symbolic notation for the convolution of the
kernels on the RHS. The retarded dissipation kernel
is defined by $\eta^r(\tau)\equiv\theta(\tau)\eta(\tau)$, and $\eta^a(\tau)
\equiv\eta^r(-\tau)$.
It can be shown that, in general, the implicit
equation (3.4) possesses two solutions for $\tilde\nu$.

In order to write Eq.(3.3b) into a compact form, observe that
the dissipation term
simulates an external potential $V_{[\xb]}(x,\tau)=2x\int_0^{\tau}
\eta(\tau-\s)\xb(\s)d\s$ as a retarded function of the label
path $\xb$. This term leads to a (label-)path dependent
contribution to the action:
$$
S_{[\xb]}[x]\equiv -\int_0^t d\tau V_{[\xb]}(x,\tau).\eqno(3.5)
$$
Furthermore, let us introduce the following norm on the space
of paths:
$$
\Vert x\Vert^2\equiv{1\over\hbar}\int_0^t d\tau \int_0^t d\s
x(\tau)\tilde\nu(\tau-\s)x(\s).\eqno(3.6)
$$
Using Eqs.(3.5) and (3.6), the compact form of
$\Phi_{[\xb]}[x]$ will be the following:
$$
\Phi_{[\xb]}[x]=exp
\left({i\over\hbar}S_{[\xb]}[x]-\Vert x-\xb\Vert^2\right).\eqno(3.7)
$$

As we have shown above, in the Caldeira-
Leggett model an exact statistical decomposition of the evolution
superoperator $J$ can explicitly be constructed.
Given the initial wave function $\psi(x,0)$, invoke
Eqs.(2.3), (3.3a) and (3.7); then introduce {\it path
dependent histories} as follows:
\begin{eqnarray}~~~~~~~~
\psi_{[\xb]}(\xfi,t)&\equiv&{1\over\sqrt {w_{[\xb]}(t)}}
\int C_{[\xb]}(\xfi,\xin)\psi(\xin,0)d\xin\nonumber\\
                         &=&{1\over\sqrt {w_{[\xb]}(t)}}\int Dx exp
\left({i\over\hbar}S[x]+{i\over\hbar}S_{[\xb]}[x]
-\Vert x-\xb\Vert^2\right)
\psi(\xin,0).~~~~~~~~~~(3.8)\nonumber\end{eqnarray}

Let us observe that this expression differs from usual
unitary Feynman integrals by the presence of the factor
$exp\left(-\Vert x-\xb\Vert^2\right)$. This factor
discards all paths from the functional integration except
for those which are close to the label path $\xb$.
Remind that the role of functional metric
specifying a distance between two paths is played by the
modified noise kernel $\tilde\nu$. We expect that a typical
history (3.8) is depicted by a wave packet propagating
along a certain label path $\xb$. The probability distribution
$w_{[\xb]}$ will mostly be concentrated on classical trajectories,
hence most likely label paths will fluctuate around classically
allowed trajectories of the central (damped) oscillator.

We have seen that to each possible path $\xb$ and to each path
dependent
history $\psi_{[\xb]}$ a certain probability can be attributed
in a consistent way. It is not at all necessary that these
consistent histories be fully decoherent.
Nevertheless, two different histories will tend to decohere:
$$
\int\psi^\ast_{[\xb_1]}(\xfi,t)\psi_{[\xb_2]}(\xfi,t)d\xfi
\rightarrow 0\eqno(3.9)
$$
if $\xb_1,\xb_2$ are two distant  paths, i.e.:
$\Vert\xb_1-\xb_2\Vert\rightarrow\infty$ [cf.Eqs.(1.2ab)].
It is seen heuristically, that decoherence is remarkable
when the distance of the two paths is
large  enough to exclude overlaps between the relevant
Feynman paths
concentrated along $\xb_1$ or $\xb_2$, respectively.
Explicit calculations are possible for the scalar
product (3.9) since the Caldeira-Leggett model possesses
exact solutions.

For pedagogical reasons, we shall
consider the very high temperature regime where the
history equations are relatively simple.

\section{Markovian Consistent histories}

It is known from, e.g., Refs.[5] that in the high temperature
limit the memory kernels $\eta,\nu$ become, in a good
approximation, local kernels.
To consider the simplest nontrivial case we assume high
temperature and {\it small velocities} $\dot x$. Then the
noise term will dominate and we shall ignore the
frictional term proportional to $\eta$. In fact we take
$\eta=0$, therefore $\nu=\tilde\nu$ holds due to Eq.(3.4).

The norm (3.6) on path space simplifies as follows:
$$
\Vert x\Vert^2
={\gamma\over\lambda_{dB}^2}\int_0^t d\tau x^2(\tau).
\eqno(4.1)
$$

The path dependent history (3.8) takes the following simple form:
$$
\psi_{[\xb]}(\xfi,t)={1\over\sqrt {w_{[\xb]}(t)}}\int Dx exp
\left({i\over\hbar}S_R[x]-\Vert x-\xb\Vert^2\right)
\psi(\xin,0).\eqno(4.2)
$$
This expression is wellknown from the theory of continuous
quantum measurements [7,6].  It is known, first of all, that the
path dependent history (4.2) is  $\psi-valued\ Markovian\ process$.
Remind the summary in Par.II, according to which
the quantity $w_{[\xb]}(t)$ in the normalizing factor yields
the probability of the given path $[\xb]$ and of the corresponding
history (4.2).
It has been shown in Refs.[8] that this process can be described
by the following Ito stochastic differential equations:
\eject
\begin{eqnarray}~~~~~~~~~~~~~~~~~~~~~~~
\dot\psi_{[\xb]}(x,\tau)&=&-{i\over\hbar}(H_R\psi_{[\xb]})(x,\tau)
-{\gamma\over2\lambda_{dB}^2}(x-<x>)^2\psi_{[\xb]}(x,\tau)+\nonumber\\
&&+(x-<x>)\psi_{[\xb]}(x,\tau)f(\tau),~~~~~~~~~~~~~~~~~~~~~~~~~~~~~~~~~~~(4.3a)\nonumber\\
\xb(\tau)&=&<x>-{\lambda_{dB}^2\over2\gamma}f(t)~~~~~~~~~~~~~~~~~~~~~~~~~
~~~~~~~~~~~~~~~~~~~~~~~~~(4.3b)\nonumber\end{eqnarray}
where $f$ is an auxiliary white noise of correlation
$$
<f(\tau)f(0)>={\gamma\over\lambda_{dB}^2}\delta(\tau).\eqno(4.3c)
$$
For the history expectation value of the coordinate
$$
<x>_{[\xb],\tau}\ \equiv\int dxx\vert\psi_{[\xb]}(x,\tau)\vert^2
\eqno(4.3d)
$$
the shorthand notation $<x>$ has been introduced.

We can see that the wave function of the path dependent history
and the path itself satisfy coupled stochastic differential
equations. From Eq.(4.3b) follows that the ordinary quantum
expectation value $<x>$ of the coordinate operator and the
label path coordinate $\xb$ will coincide in stochastic mean.

Incidentally, it is perhaps instructive to write Eq.(4.3a) into
an equivalent Ito form for the pure state statistical
operator
$P_{[\xb]}(x,\xp)\equiv\psi_{[\xb]}(x)\psi^\dagger_{[\xb]}(\xp)$:
\begin{eqnarray}~~~~~~~~~~~~
\dot P_{[\xb]}(x,\xp,\tau)&=&-{i\over\hbar}[H_R,P_{[\xb]}](x,\xp,\tau)
-{\gamma\over2\lambda_{dB}^2}(x-\xp)^2P_{[\xb]}(x,\xp,\tau)\nonumber\\
&&~~~~~~+(x+\xp-2<x>)P_{[\xb]}(x,\xp,\tau)f(\tau).
~~~~~~~~~~~~~~~~~~~~~~~~(4.4)\nonumber\end{eqnarray}
Taking the stochastic average of both sides, the nonlinear term
cancels and we obtain the wellknown linear Markovian master
equation (A.11).

The Markovian history equations (4.3ab) can be solved exactly
in the long time limit. In Refs.[8,9] the following result was
obtained for the special case,
when the renormalized central oscillator
is just a free particle, i.e. $\Omega_R=0$.
The shape of the wave function becomes stabilized at the Gaussian
shape, i.e.
$$
\psi_(x,\tau)\sim (2\pi\s^2)^{-1/4}
exp\left(ix<p>-{1-i\over4\s^2}(x-<x>)^2\right)
\eqno(4.5a)
$$
of width $\s=(\hbar/2)^{3/4}(\gamma k_BT)^{-1/4}M^{-1/2}$,
while the quantum expectation values
\hbox{$<x>,<p>$} of the coordinate
and momentum, resp., satisfy the following stochastic equations:
\begin{eqnarray}~~~~~~~~~~~~~~~~~~~~~~~~~~~~~~~~~~~~~~~~~
{d<x>\over d\tau}&=&{<p>\over M}+2\sigma^2f,\nonumber\\
         {d<p>\over d\tau}&=&\hbar f.~~~~~~~~~~~~~~~~~~~~~~~~~~~~~~~~~~~~~~~~~~~~~~~~~~
(4.5b)\nonumber\end{eqnarray}

We note that similar quality of results would be obtained
from more general Ito differential equations (cf.Ref.[10]),
had we retained the dissipation term proportional to $\nu$
(A.9b).

\section{Summary}

Starting from the conservative statistical interpretation of mixed
quantum states, we have proposed a family of quantum histories
possessing consistent probability distribution. The proposal
has been realized for the Caldeira-Leggett model of Brownian
motion. Our
history expansion is exact and needs {\it no particular tuning
of decoherence and coarse graining}. In the Markovian regime,
we have obtained analytic localized solutions for the Brownian
particle's wave function.

\section*{A. Coarse graining in the Caldeira-Leggett model [5]}

Our central system is a harmonic oscillator of mass M and
frequency $\Omega$, with action
$$
S[x]={M\over2}\int_0^t d\tau(\xd^2-\Omega^2 x^2).\eqno(A.1)
$$
The initial quantum state will be denoted by $\rho(\xin,\xpi,0)$ .

Consider a reservoir modeled by a set of harmonic oscillators
with masses $m_n$ and with frequencies $\omega_n$. Its action is:
$$
S_{res}[q]=
\sum_n {m_n\over2}\int_0^t d\tau(\qd_n^2-\omega_n^2 q_n^2).\eqno(A.2)
$$
At $t=0$ the state
of the reservoir is thermal equilibrium state at some
temperature $T$. Consider a certain linear combination
$Q=\sum_nc_nq_n$ of the reservoir coordinates. Introduce
the complex
correlation function of the Heisenberg operators $Q(\tau)$:
$$
\nu(\tau)+i\eta(\tau)\equiv{1\over\hbar}<Q(\tau)Q(0)>_T\eqno(A.3)
$$
where $<...>_T$ stands for the expectation value taken in the
thermal equilibrium state of the reservoir. The real and imaginary
parts $\nu,\eta$ are called the noise (or fluctuation) and the
dissipation kernels, respectively.

For $t>0$, a linear coupling is
introduced between the central oscillator and the reservoir,
represented by the action \footnote{The printed version contained the typo
$qQ$ instead of the correct $xQ$.}
$$
S_{int}[x,q]=-\int_0^td\tau xQ.\eqno(A.4)
$$

The usual coarse graining of the above system consists
of tracing out the variables of the reservoir. Then
the statistical operator of the central oscillator obeys to
linear evolution equation:
$$\rho(\xfi,\xpf,t)=
  \int J(\xfi,\xpf,\xin,\xpi,t)
  \rho(\xin,\xpi,0)d\xin d\xpi.\eqno(A.5)
$$
The superoperator $J$ takes the following general form:
$$
J(\xfi,\xpf,\xin,\xpi,t)=\int Dx \int D\xp
exp\left({i\over\hbar}S[x]-{i\over\hbar}S[\xp]\right)F[x,\xp]
\eqno(A.6)
$$
with the decoherence functional $F$ defined by
\begin{eqnarray}~~~~~~~~~~~~~~~~~~
F[x,\xp]=exp\biggl(&&\!\!\!\!\!\!\!\!-{i\over\hbar}\int_0^t d\tau
                                           \int_0^\tau d\tp
         [x(\tau)-\xp(\tau)]\eta(\tau-\tp)[x(\tp)+\xp(\tp)]\nonumber\\
                   &&\!\!\!\!\!\!\!\!-{1\over2\hbar}\int_0^t d\tau
                                            \int_0^t d\tp
         [x(\tau)-\xp(\tau)]\nu (\tau-\tp)[x(\tp)-\xp(\tp)]\biggr).
~~~(A.7)\nonumber\end{eqnarray}
For the Caldeira-Leggett reservoir, the  noise and dissipation
kernels have the following particular forms, respectively:
$$
\nu(\tau)=\ {\gamma M\over \pi}\int_0^{\omega_{max}}d\omega
        \omega\coth{\hbar\omega\over 2k_BT}\cos(\omega\tau)
\eqno(A.8a)
$$
$$
\eta(\tau)=-{\gamma M\over \pi}\int_0^{\omega_{max}}d\omega
          \omega\sin(\omega\tau).
\eqno(A.8b)
$$

For high temperatures, Markovian approximation can be applied
to the noise and dissipation kernels:
$$
\nu(\tau)={\gamma\hbar\over \lambda_{dB}^2}\delta(\tau),\eqno(A.9a)
$$
$$
\eta(\tau)=\gamma M\delta^\prime(\tau)\eqno(A.9b)
$$
where $\lambda_{dB}=\hbar/\sqrt{2Mk_BT}$ is the thermal
deBroglie length.
In addition, the frequency $\Omega$ of the central oscillator
must be replaced by its renormalized value, defined by
$\Omega_R^2=\Omega^2-2\gamma\Omega/\pi$.
Consequently, the
action $S$ on the RHS. of Eq.(A.6) will be replaced by the
renormalized action:
$$
S_R[x]={M\over2}\int_0^t d\tau(\xd^2-\Omega_R^2x^2).\eqno(A.10)
$$

In Markovian approximation the evolution superoperator (A.6)
becomes local in time. Hence the evolution equation (A.5) can
equivalently be written in form of a linear differential
equation. We do not quote the general result but a simplified
version valid for small velocities $\xd$:
$$
\dot\rho(x,\xp,\tau)
=-{i\over\hbar}[H_R,\rho](x,\xp,\tau)
-{\gamma\over2\lambda_{dB}^2}(x-\xp)^2
\rho(x,\xp,\tau).\eqno(A.11)
$$
The general master equation contains additionally a certain
dissipation term proportional to the momentum p, and a further
fluctuation term [11] as well.

This work was supported by the
Hungarian Scientific Research Fund under Grant No 1822/1991.

\end{document}